# Statistics of Hotspots in Radio Galaxies

**Pedro Abel de la Rosa Valdés**
Universidad Autónoma de Zacatecas
Calzada Solidaridad esq. Paseo La Bufa s/n
C.P. 98060, Zacatecas, Zac., Mexico
pedrodelarosa993@gmail.com

**Heinz Andernach**
Universidad de Guanajuato
Departamento de Astronomía, DCNE
C.P. 36023, Guanajuato, Gto., Mexico
heinz@astro.ugto.mx

***Resumen*** — Con base en imágenes del rastreo del cielo en radio FIRST a 1.4 GHz, se compilaron las posiciones de unos 5200 *hotspots* en ~2870 radiogalaxias y cuásares, cuyos tamaños lineales y radioluminosidades fueron derivados de los *redshifts* de los huéspedes. Para una submuestra de ~2100 radiofuentes con un *hotspot* en cada uno de sus lóbulos, se estudió su geometría en términos de asimetría y grado de torsión. Se comprobó la (débil) tendencia conocida que el lóbulo (aquí *hotspot*) más brillante se ubica más cerca del huésped que el *hotspot* menos brillante. La mediana del ángulo de torsión entre los dos brazos de las radiofuentes es de 4.8° con una diferencia significativa en la distribución entre las 627 fuentes con huéspedes cuásares y las 1501 galaxias.

***Palabras clave*** — Rastreos en radio, radiogalaxias, hotspots.

***Abstract*** — Based on images from the FIRST survey of the radio sky at 1.4 GHz, the positions of ~5200 hotspots in ~2870 radio galaxies and quasars were compiled, and linear sizes and radio luminosities were derived from the hosts redshifts. For a subsample of ~2100 radio sources with exactly one hotspot in each of the two opposite lobes, their geometry in terms of asymmetry and bending was studied. The known (weak) tendency for the brighter lobe (here hotspot) to lie closer to the host than the fainter one, is confirmed. The median bending angle between the two arms of radio sources is 4.8° with a significant difference in the distribution between the 627 quasar hosts and the 1501 galaxies.

***Keywords*** — radio surveys, radio galaxies, hotspots.

## I. INTRODUCTION

Radio galaxies (RGs, including quasar hosts unless otherwise noted) show an excess of radio continuum emission over normal galaxies, and their radio structure consists of a radio nucleus (if present) at the core of the host galaxy from where two jets in opposite directions carry relativistic particles usually far beyond the outskirts of the optical galaxy to form outer radio lobes of varied shapes. Many of these feature "hotspots", a term originally coined by Gull & Northover (1973) for the most compact emission regions in the lobes of Cygnus A. Hotspots are likely the places where the relativistic particles transported from the galaxy nucleus are reaccelerated, since for most RGs the synchrotron radiation lifetime of these particles is shorter than their travel time from the nucleus to the outer lobes.

While radio hotspots are ubiquitous in edge-brightened (or FR II type, Fanaroff & Riley 1974) radio galaxies, of those ~100 hotspots in the literature studied outside the radio window, only two-thirds were detected in X-rays, about one-third in the infrared (IR) and only ~16% in the optical (see e.g. Hardcastle et al. 2004, Werner et al. 2012). Valdés Ochoa (2019) has shown that the average radio-to-IR spectral slope of hotspots individually IR-undetected can be inferred from stacking many hundreds of mid-IR images drawn from the WISE mission (Cutri et al. 2013) and centered on radio-detected hotspots.

Hotspots have also been used to characterize the symmetry of FR II RGs and to infer the hotspot advance speed, with sometimes contradicting results, partly because of too small samples available at the time. Here we aim to expand the number of catalogued hotspots. We used standard cosmology with $H_0$=70 km s$^{-1}$ Mpc$^{-1}$, $\Omega_m=0.3$ and $\Omega_\Lambda=0.7$ to derive linear sizes and luminosities.

## II. MATERIALS AND METHODS

For the construction of this new sample of hotspots we used a compilation of extended RGs maintained by one of us (see e.g. Andernach 2018), and which includes for each object the position and brightness of its optical or IR host, the measured or photometric redshift, and the largest angular size (LAS) and largest linear size (LLS) of its radio emission. In a summer research project in 2016 (see Valdés Ochoa 2019 for the final result) radio images from the "Faint Images of the Radio Sky" (FIRST, Helfand et al. 2015) survey at 1.4 GHz and 5.4" angular resolution were extracted and used to construct a list of ~2680 hotspots in ~1530 RGs. Since 2016, many more FRII-type radio galaxies were compiled by H.A. such that in a 2018 summer internship, D.E. Monjardin Ward established a further list of ~1150 hotspots in 605 radio galaxies. This list was revised by us in the present work, discarding ~5% of the originally chosen hotspots and adding another ~5% that had been missed. Since 2018, a further ~750 radio galaxies were compiled that appeared to have hotspots, and their FIRST cutouts were used here to construct a further list of ~1440 hotspots in 737 RGs.

To aid in recognizing the radio source structure we also used the lower-resolution radio surveys NVSS (Condon et al. 1998), and TGSS-ADR1 (Intema et al. 2017), and occasionally consulted images from the ongoing "VLA Sky Survey" at 2-4 GHz with 2.5" angular resolution (VLASS, Lacy et al. 2019) to assess the reality of a hotspot, but taking the data from FIRST only. Images of the same source were displayed with ObitView (http://www.cv.nrao.edu/~bcotton/Obit.html) in order to register the exact host positions with its "Gaussfit" option, as well as with ds9 (http://ds9.si.edu) and Aladin (Bonnarel et al. 2000), which made it easier to distinguish the hotspots from more diffuse emission.

Here we combine the above-mentioned three samples to obtain a list of over 5200 hotspots in 2869 RGs for a statistical analysis of hotspot parameters and the degree of symmetry of classical double radio sources. These RGs span a wide range of redshifts with median z=0.53 with quartiles at 0.35 and 0.73, and 373 RGs lie at redshifts above 1. The median LAS is 1.65', with quartiles at 1.1' and 2.5'. Their LLS ranges from 50 kpc to almost 5 Mpc, with a median of 580 kpc and quartiles at 400 and 820 kpc. Spectroscopic redshifts are available for ~50% of the sample and for most others reliable photometric ones were culled from references like DiPompeo et al. (2015), Bilicki et al. (2016), and others.

The hotspot positions from ObitView were then cross-matched with the FIRST source catalog using the VizieR catalog browser (Ochsenbein et al. 2000) choosing either the closest source within 12", or the farthest one from the host as the hotspot. Less than one percent of the hotspots had no counterpart in the FIRST catalog, either for having a peak brightness of < 1 mJy/beam or being located close to the FIRST survey boundary, and were discarded, leaving 5236 hotspots in 2869 RGs. For each FIRST source its integrated flux (from the FIRST catalogue) was combined with the redshift of the host to find the hotspots 1.4-GHz radio luminosity, $logP_{1.4}$, and their deconvolved major axes were used to derive linear hotspot sizes, $LS_{hs}$. Angular and linear lengths and position angles, $PA_{arm}$, of the vectors from host to hotspot (called the "arm" of a radio galaxy) were computed, as well as the acute angle, $PA_{dif}$, between the hotspot major radio axis ($PA_{hs}$) and $PA_{arm}$.

### III. RESULTS AND ANALYSIS

Of the 2869 RGs with any hotspot, we found 21% to have one hotspot, 75% with two, 3.6% with three, and 0.3% with four hotspots. Note that these numbers are biased, as they are based on a selection of RGs which showed at least one hotspot at first sight on a FIRST image. Experience showed (Valdés Ochoa 2019) that about 20−25% of all non-FR I sources do not have any hotspot. We further found that 178 (3.4 %) of our hotspots were labelled with a sidelobe probability p(S) > 7 % in the FIRST catalog. Since 2.9 % of all 946,432 FIRST sources have a p(S) > 7 %, we anticipate that ~800 further such sources will turn out to be hotspots rather than artefacts.

The linear size distribution of 5236 hotspots has a median of 34 kpc and is shown in figure 1. The long tail of larger sizes is likely due to hotspots embedded in the diffuse emission of the surrounding lobe. The decimal log of the 1.4-GHz luminosity (in W/Hz) of these hotspots has a median of 25.26 and its distribution is shown in figure 2.

Figure 3 shows the distribution of the misalignment angle between the host-to-hotspot direction and the orientation of the hotspot major radio axis. While the majority of hotspots appear to be aligned with the "arm" in which they reside (51% within 20° vs. 9.6% beyond 70°), a plot of the hotspot linear sizes vs. the misalignment angle shows that the aligned hotspots are on average larger, suggesting that part of this trend is due to hotspots embedded in diffuse lobe emission, and that those hotspots close to perpendicular to the arm axis may be bow shocks at the working surface between jet and intergalactic medium.

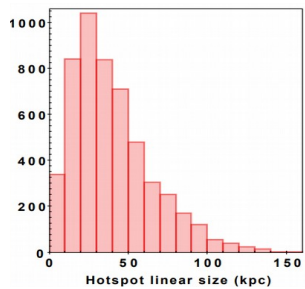

**Fig. 1.** Hotspot linear size distribution.

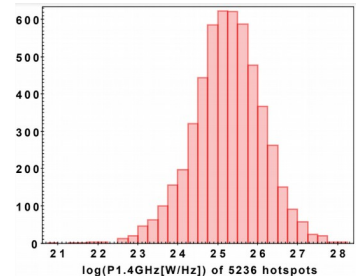

**Fig. 2.** Distribution of 1.4-GHz luminosity for 5236 hotspots

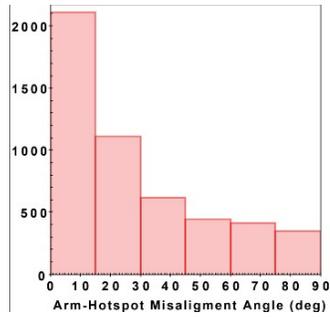

**Fig. 3.** Distribution of misalignment angle between hotspot and arm orientation.

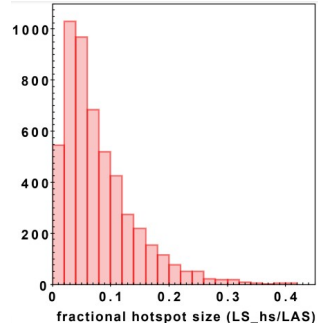

**Fig. 4.** Distribution of fractional hotspot sizes for 5192 hostspots.

Figure 4 displays the distribution of the "fractional hotspot size", the ratio FHS = $LS_{hs}$/LAS, with a median of 0.062 for those 5192 hotspots with non-zero major-axis values in the FIRST catalogue. For 919 hotspots with hosts at z<0.3 the median FHS decreases to 0.05. This is five times larger than the FHS found by Hardcastle et al. (1998) since their value of ~0.01 was based on observations of up to 10 times higher angular resolution than FIRST. Indeed, those 83 sources with LAS>3' (>33 FIRST beams) have $FHS_{med}$=0.028.

For a subsample of 2128 RGs with one hotspot on each side, we assumed these to be the "ends" of the sources. To check this assumption we calculated the ratio, SSL, of the sum of the two hotspot separations from the host divided by the LAS of the RG. The 31 RGs with SSL<0.5 are mostly of the "restarted" (or double-double), X-shaped or precessing type, and very few of these may be wide-angle tailed (WAT) sources with their radio knots misidentified as hotspots. Another 11 RGs with SSL>1.2 show radio emission beyond their hotspots. Excluding these few sources will not alter the following results.

We calculated the "armlength ratio" (ALR) as the distance from the host object to the stronger hotspot divided by the distance to the fainter hotspot, as well as the "flux ratio" (FLR) between the integrated flux density (as listed in the FIRST catalog) of the hotspot more distant from the host to that of the closer one, as well as the acute angle between the two position angles from the host to each of the two hotspots, the "misalignment" or bending angle, BA, of the radio galaxy or quasar.

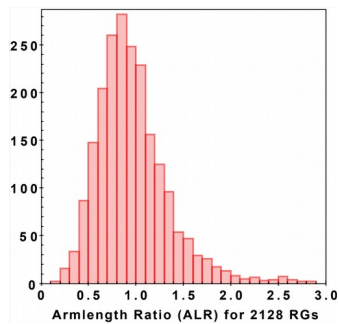

**Fig. 5.** Distribution of armlength ratio

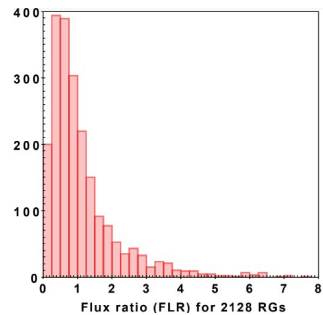

**Fig. 6.** Distribution of flux ratio

Figures 5 and 6 show the distributions of ALR and FLR, except for a few outliers with ALR>3 and FLR>8. Their medians are 0.91 and 0.815, respectively. Both figures confirm the known trend that the stronger lobe (here hotspot) tends to lie closer to the host.

The 2128 RGs have a median BA of 4.8°, with little difference between the 1501 galaxies (4.6°) and 627 quasars (5.1°), but distributed differently at 99.4% confidence.

## IV. CONCLUSIONS AND FUTURE WORK

We composed an unprecedented sample of over 5200 hotspots in 2869 radio galaxies and quasars covering a large range of redshifts, radio luminosities and projected linear radio sizes. Based on source sizes from the FIRST catalog, the median size of hotspots is 34 kpc, compared to 20 kpc for the 65 hotspots in Hardcastle et al. (2004), which was based on observations of much higher angular resolution.

For a sample of 2128 radio sources with one hotspot per side, we confirm the known fact that on average the brighter lobe (here hotspot) is closer to the host than the fainter one. The median projected bending angle between the two arms of radio sources is ~5°, but their distribution is much broader for the 627 quasars than for the 1501 galaxy hosts.

Longair & Riley (1979) and Banhatti (1980) have proposed to use what they call the "separation quotient", Q, and "fractional separation difference", X, in order to estimate the advance speed of the hotspots. While we do not confirm the finding of Best et al. (1995) that the X distribution of quasars has a minimum near zero, we do find that X values of quasars have a broader distribution than that of galaxies at 100.00% confidence, which is consistent with the expectation that quasar jets make a smaller angle with our line of sight.

Our new hotspot sample will be ideal to be matched with modern optical, IR, and X-ray surveys, in order to either find their SEDs or discard some of them for being superposed on the radio galaxies. For the vast majority of hotspots undetected in these non-radio bands, a stacking analysis could shed light on their average SED. The ongoing VLASS survey will allow to determine the positions of hotspots in many times the current number of radio galaxies, moreover at two times better angular resolution than FIRST, allowing to better separate real hotspots from mere brightness peaks in the diffuse lobes, which certainly constitute a (small) fraction of the hotspots used in the present work.

### V. ACKNOWLEDGEMENTS

We thank Isabel Valdés Ochoa and Douglas Monjardin Ward for assembling ~60% of the hotspot sample used here. A significant number of RGs in our sample was found by volunteers of Radio Galaxy Zoo (http://rgzauthors.galaxyzoo.org). We are grateful to the VLASS team at NRAO for a timely delivery of Quick Look images, and to F.J. Peralta for a python script to extract VLASS cutouts. W. Cotton (NRAO) provided a revised script to download NVSS images. H. A. benefited from Univ. Guanajuato grant CIIC 218/2019.